# Size-dependent spatial magnetization profile of Manganese–Zinc ferrite $Mn_{0.2}Zn_{0.2}Fe_{2.6}O_4$ nanoparticles

Mathias Bersweiler,[1] Philipp Bender,[1] Laura G. Vivas,[1]
Martin Albino,[2] Michele Petrecca,[2,3]
Sebastian Mühlbauer,[4]
Sergey Erokhin,[5] Dmitry Berkov,[5]
Claudio Sangregorio,[2,3] and Andreas Michels[1]

[1]*Physics and Materials Science Research Unit, University of Luxembourg, 162A Avenue de la Faïencerie, L-1511 Luxembourg, Grand Duchy of Luxembourg*
[2]*Università degli Studi di Firenze, Dipartimento di Chimica "U. Schiff", Via della Lastruccia 3, 50019 Sesto Fiorentino, Italy*
[3]*ICCOM-CNR via Madonna del Piano 10, 50019 Sesto Fiorentino, Italy*
[4]*Heinz Maier-Leibnitz Zentrum (MLZ), Technische Universität München, D-85748 Garching, Germany*
[5]*General Numerics Research Lab, An der Leite 3B, D-07749 Jena, Germany*

We report the results of an unpolarized small-angle neutron scattering (SANS) study on Mn-Zn ferrite (MZFO) magnetic nanoparticles with the aim to elucidate the interplay between their particle size and the magnetization configuration. We study different samples of single-crystalline MZFO nanoparticles with average diameters ranging between 8 to 80 nm, and demonstrate that the smallest particles are homogeneously magnetized. However, with increasing nanoparticle size, we observe the transition from a uniform to a nonuniform magnetization state. Field-dependent results for the correlation function confirm that the internal spin disorder is suppressed with increasing field strength. The experimental SANS data are supported by the results of micromagnetic simulations, which confirm an increasing inhomogeneity of the magnetization profile of the nanoparticle with increasing size. The results presented demonstrate the unique ability of SANS to detect even very small deviations of the magnetization state from the homogeneous one.

## I. INTRODUCTION

The Manganese-Zinc ferrite (MZFO) material system possesses favorable physical properties such as high magnetic permeability, reasonable saturation magnetization combined with low eddy current losses, high electrical resistivity as well as a good flexibility and chemical stability. These features render MZFO a very promising candidate for many technological and biomedical applications, e.g., as magnetic reading heads [1], constituents of temperature-sensitive ferrofluids [2], microwave absorbers [3], inductors [4], drug delivery [5,6], and MRI contrast enhancing agents [7]. A problem arises because the macroscopic magnetic properties of MZFO are strongly dependent e.g. on their chemical composition [8–10], the synthesis methods [11,12], and on the distribution of cations between interstitial tetrahedral and octahedral sites [9,13,14]. Moreover, even for the same chemical composition, the magnetic properties may sensitively depend on the MZFO particle size [8,10,15,16].

Previous studies on MZFO nanoparticles along these lines using conventional magnetometry have reported a transition from single- to multi-domain structure for critical sizes between about 20-40 nm [8,15]. In the present work, we employ magnetic-field-dependent unpolarized small-angle neutron scattering (SANS) to obtain *mesoscopic* information on the magnetization profile within MZFO nanoparticles of different sizes. Magnetic SANS provides volume-averaged information about variations of the magnetization vector field on a nanometer length scale of ∼ 1 – 100 nm (see Refs. [17,18] for reviews).

The SANS technique has been used in several other studies to investigate intra and interparticle magnetic moment correlations in various nanoparticle systems; for instance, SANS was applied to study interacting nanoparticle ensembles [19,20], including ordered arrays of nanowires [21,22], it was employed to reveal the domain orientation in nanocrystalline soft magnets [23], or to investigate the response of magnetic colloids [24–26] and ferrofluids [27–29] to external fields. In Refs. [20,30–32] the SANS method has been utilized to disclose the *intraparticle* magnetization profile on different magnetic nanoparticle systems. These studies indicate the presence of spin disorder and canting, particularly at the nanoparticle surface. A nonuniform spin texture obviously affects the macroscopic magnetic properties, and hence the application potential. Here, we also use magnetic SANS to disclose the magnetization profile, however, in contrast to the previous works we focus our analysis on *model-independent* approaches. Additionally, we use large-scale micromagnetic continuum simulations to support our findings and to disclose the delicate interplay between particle size and magnetization profile within MZFO nanoparticles.

The article is organized as follows: In Section II, we discuss the nanoparticle synthesis, the characterization methods, and the details of the SANS experiment. In Section III, we summarize briefly the expressions for the unpolarized SANS cross section, the intensity ratio, and the correlation function. Section IV presents and discusses the experimental results of the characterization of the samples by X-ray fluorescence spectrometry, X-ray diffraction, transmission electron microscopy, magnetometry, and in particular the

SANS measurements; a paragraph on the micromagnetic simulation results completes this section. Section V summarizes the main findings of this paper.

## II. EXPERIMENTAL

$Mn_{0.2}Zn_{0.2}Fe_{2.6}O_4$ nanoparticles covered with a monolayer of oleic acid (capping agent) were synthesized by co-precipitation from aqueous solutions and by thermal decomposition of iron and manganese acetylacetonates in high-boiling solvent (benzyl ether) in the presence of surfactants and of $ZnCl_2$ (see Appendix A and B for details on the nanoparticle synthesis). In the following, the particles will be labeled as MZFO-x, where x denotes their average particle size.

The chemical composition of the nanoparticles was determined by a Rigaku ZSX Primus II X-ray fluorescence spectrometer (XRF), equipped with a Rh Kα radiation source and a wavelength dispersive detector. The average crystallite size and the structural properties of the nanoparticles were estimated by transmission electron microscopy (TEM), using a CM12 Philips microscope with a $LaB_6$ filament operating at 100 kV, and by X-ray diffraction (XRD), using a Bruker New D8 ADVANCE ECO diffractometer with Cu Kα radiation. The amount of organic layer was estimated by CHN analysis, using a CHN-S Flash E1112 Thermofinnigan. The magnetic analysis at room temperature was performed on tightly packed powder samples using a Quantum Design MPMS superconducting quantum interference device (SQUID) magnetometer.

For the SANS experiments, the nanoparticles were pressed into circular pellets with a diameter of 8 mm and a thickness of $1.3 \pm 0.1$ mm. The neutron experiments were performed at the instrument SANS-1 [33] at the Heinz Maier-Leibnitz Zentrum (MLZ), Garching, Germany. The measurements were done using an unpolarized incident neutron beam with a mean wavelength of $\lambda = 4.51$ Å and a wavelength broadening of $\Delta\lambda/\lambda = 10$ % (FWHM). All the measurements were conducted at room temperature and within a $q$-range of about $0.06$ nm$^{-1} \leq q \leq 3.0$ nm$^{-1}$. A magnetic field $H_0$ was applied perpendicular to the incident neutron beam ($H_0 \perp k_0$). The experimental setup used for these experiments is sketched in Fig. 1. Neutron data were recorded by increasing the applied magnetic field from 0 T to 4 T following the magnetization curve. The neutron-data reduction (correction for background and empty cell scattering, sample transmission, detector efficiency, and water calibration) was carried out using the GRASP software package [34].

## III. SANS CROSS SECTION, INTENSITY RATIO, AND CORRELATION FUNCTION

### A. Elastic unpolarized SANS cross section

As detailed in Refs. [17,18], when the applied magnetic field $H_0$ is perpendicular to the incident neutron beam ($H_0 \perp k_0$), the elastic nuclear and magnetic unpolarized SANS cross section $d\Sigma/d\Omega$ at momentum-transfer vector $q$ can be written as:

$$\frac{d\Sigma}{d\Omega}(q) = \frac{8\pi^3}{V} b_H^2 \left( b_H^{-2}|\tilde{N}|^2 + |\tilde{M}_x|^2 + |\tilde{M}_y|^2 \cos^2(\theta) + |\tilde{M}_z|^2 \sin^2(\theta) \right. \\ \left. - (\tilde{M}_y \tilde{M}_z^* + \tilde{M}_y^* \tilde{M}_z) \sin(\theta) \cos(\theta) \right), \quad (1)$$

where $V$ is the scattering volume, $b_H = 2.91 \times 10^8$ A$^{-1}$m$^{-1}$ relates the atomic magnetic moment to the atomic magnetic scattering length, $\tilde{N}(q)$ and $\tilde{M}(q) = [\tilde{M}_x(q), \tilde{M}_y(q), \tilde{M}_z(q)]$ represent the Fourier transforms of the nuclear scattering length density $N(r)$ and of the magnetization vector field $M(r)$, respectively, $\theta$ specifies the angle between $H_0$ and $q$ (see Fig. 1), and the asterisks "*" denote the complex conjugated quantity. Generally, the Fourier components $\tilde{M}_{x,y,z}$ depend on both the magnitude and the orientation of the scattering (wave) vector $q$. This dependence is influenced by the applied magnetic field, the various intra and interparticle magnetic interactions, and by the particle size and shape. It is also worth emphasizing that in the small-angle approximation (scattering angle $\psi \ll 1$) only correlations in the plane perpendicular to the incoming neutron beam are probed (compare Fig. 1); this means that the above Fourier components are to be evaluated at $q_x \cong 0$.

### B. SANS intensity ratio

Deviations from the uniform magnetization state in nanoparticle systems had already become evident in the early SANS study by Ernst, Schelten, and Schmatz [35]. These authors investigated the transition from single to multi-domain configurations of Co precipitates in a Cu single crystal and analyzed the following ratio $\alpha(q)$ of SANS cross sections ($H_0 \perp k_0$) [35]:

$$\alpha(q) = \frac{\left.\frac{d\Sigma}{d\Omega}(q)\right|_{H_0 = 0\,\text{T}}}{\left.\frac{d\Sigma}{d\Omega}(q)\right|_{q \parallel H_0 \to \infty}} = \frac{\left[\frac{d\Sigma}{d\Omega}_{\text{nuc}}(q) + \frac{d\Sigma}{d\Omega}_{\text{mag}}(q)\right]\bigg|_{H_0 = 0\,\text{T}}}{\left.\frac{d\Sigma}{d\Omega}_{\text{nuc}}(q)\right|_{q \parallel H_0 \to \infty}}. \quad (2)$$

The total unpolarized SANS cross section $d\Sigma/d\Omega$ at zero applied magnetic field equals the sum of nuclear and magnetic contributions, while the cross section at a saturating field $H_0$ applied parallel to the scattering vector $q$ yields (for $k_0 \perp H_0$) the purely nuclear SANS cross section $d\Sigma_{\text{nuc}}/d\Omega$. We emphasize that the interpretation of $\alpha(q)$ is highly nontrivial, since it depends on a number of both structural and magnetic parameters; for instance, on the particle volume fraction, at high packing densities also on the shape and size distribution of the particles, and not the least on the internal spin structure of the nanoparticles, which depends e.g. on the particle size and the applied field, but also on the strength of the magnetodipolar interaction between the particles.

Consider the special case of a *dilute* assembly of randomly-oriented single-domain particles: if for $H_0 = 0$, the magnetizations of the particles are randomly oriented, then the two-dimensional $d\Sigma/d\Omega$ is isotropic, whereas it exhibits the well-known $\sin^2\theta$ angular anisotropy for the saturated case, $k_0 \perp H_0$, and for a not too strong nuclear signal

[compare Eq. (1)]. For this particular situation, the ratio $\alpha$ depends only on the magnitude $q$ of the scattering vector:

$$\alpha(q) = \frac{\left[\frac{d\Sigma}{d\Omega}_{\text{nuc}}(\boldsymbol{q}) + \frac{d\Sigma}{d\Omega}_{\text{mag}}(\boldsymbol{q})\right]\bigg|_{\boldsymbol{q} \parallel \boldsymbol{H}_0 = 0\,\text{T}}}{\frac{d\Sigma}{d\Omega}_{\text{nuc}}(\boldsymbol{q})\bigg|_{\boldsymbol{q} \parallel \boldsymbol{H}_0 \to \infty}} = 1 + \frac{\frac{d\Sigma}{d\Omega}_{\text{mag}}(q)}{\frac{d\Sigma}{d\Omega}_{\text{nuc}}(q)}, \quad (3)$$

where the isotropic zero-field SANS cross section has also been averaged for $\boldsymbol{q} \parallel \boldsymbol{H}_0$. By contrast, for a globally anisotropic microstructure, e.g., for oriented shape-anisotropic particles or for a system exhibiting a large remanence [33], the $\alpha$-ratio may depend on the orientation of $\boldsymbol{q}$. Moreover, if in the dilute ensemble of randomly-oriented single-domain particles the chemical (nuclear) and magnetic particle sizes coincide, then Eq. (3) simplifies to the $q$-independent value:

$$\alpha_{\text{calc}} = 1 + \frac{2}{3}\left(\frac{\varrho_{\text{mag}}}{\Delta\varrho_{\text{nuc}}}\right)^2, \quad (4)$$

where $\Delta\varrho_{\text{nuc}}$ is the difference between the nuclear scattering length densities of the nanoparticles and the matrix, and $\varrho_{\text{mag}} = b_H M_S^{MZFO}$ is the magnetic scattering length density of the MZFO nanoparticles. The factor 2/3 in Eq. (4) results from an orientational average of the $\sin^2(\theta)$ factor in Eq. (1) in the remanent state (assuming the absence of other magnetic scattering contributions in line with the assumption of the presence of only single-domain particles). Under the above assumptions, deviations from the constant value given by Eq. (4) may indicate the presence of intra-particle spin disorder.

### C. Correlation function

To obtain real-space information about the magnetic microstructure, we have computed the following correlation function [36–39]:

$$p(r) = r^2 \int_0^\infty I(q) j_0(qr) q^2 dq, \quad (5)$$

where $j_0(x) = \sin(x)/x$ denotes the spherical Bessel function of zero order, and $I(q)$ represents the azimuthally-averaged magnetic SANS cross section. In nuclear SANS and small-angle X-ray scattering $p(r)$ is known as the pair-distance distribution function, which provides information on the particle size and shape, and on the presence of interparticle interactions; for magnetic systems it may also indicate the presence of intraparticle spin disorder.

## IV. RESULTS AND DISCUSSION

### A. Structural and magnetic pre-characterization

XRF analyses confirmed that the synthesized nanoparticle samples all have a similar composition, i.e., $Mn_{0.2}Zn_{0.2}Fe_{2.6}O_4$ (see Table 1). XRD results for the nanoparticle powders are shown in Fig. 2(a). All the diffraction peaks observed can be well indexed with the $AB_2O_4$ spinel structure, indicating a pure cubic phase of $Mn_{0.2}Zn_{0.2}Fe_{2.6}O_4$. Moreover, impurity peaks or secondary phases are not observed in our XRD pattern, which confirms the high quality of the nanoparticles synthetized by co-precipitation and thermal decomposition. The structural parameters were determined by the method of the fundamental parameter approach (FDA) implemented in the TOPAS software, considering the cubic space group $Fd\bar{3}m$. The average crystallite sizes are reported in Table 1. The lattice parameter $a$ varies in the range from 0.8407(2) to 0.8421(1) nm, as expected for doped Mn-Zn ferrite nanoparticles [9].

TEM images of the nanoparticles are displayed in Fig. 2(b) and the average particle sizes are listed in Table 1. It should be emphasized that the small nanoparticles look spherical, whereas the larger nanoparticles seem to have a faceted cubic structure. This morphology evolution is the result of the interplay between surface tension and preferential growth along the <100> directions [40]. For all samples, the average particle size determined by TEM is nearly identical to the XRD crystallite size, suggesting that the nanoparticles are single crystals. The CHN analysis indicates that the relative amount of surfactant decreases with the nanoparticle surface-to-volume ratio, from 11.2 % for MZFO-8 to 1.1 % for MZFO-80. For all the samples, this corresponds approximately to a monolayer of surfactant, as evaluated by assuming that each ligand molecule occupies a surface area of 0.5 nm$^2$ [19,41]. Figure 3 shows a typical scanning electron microscopy (SEM) image of a MZFO sample after the powder has been pressed into a circular pellet; this microstructure is characteristic of the SANS samples in our study.

The normalized room-temperature magnetization curves $M(H)$ of the nanoparticle powders are shown in Fig. 4(a) and in Fig. 4(b), respectively. From these curves, we determined the saturation and remanent magnetizations ($M_S$ and $M_R$ respectively) and the coercive field $H_C$ (see Table 1). The $M(H)$ curve of MZFO-8 shows no hysteresis, indicating superparamagnetic behavior. However, for larger particle sizes, the $M(H)$ curves start to open up and an increase of $M_S$, $M_R$, and $H_C$ is observed.

From the $M(H)$ curve of MZFO-8, we have extracted the underlying effective moment distribution $P_V(\mu)$ using the approach outlined in Bender *et al.* [42] [see Fig. 4(c)], where a Langevin-type magnetization behavior is assumed. The obtained distribution exhibits one main peak at $\sim 10^{-19}$ Am$^2$ and additional contributions in the low-moment range. We surmise that the main peak corresponds to the distribution of the individual particle moments $\mu_i = M_S V_i$ of the whole ensemble (where $V_i$ is the particle volume), and that the low-moment contributions can be attributed to dipolar interactions within the ensemble, similar as in Bender *et al.* [42]. As shown in Fig. 4(c), the main peak can be well adjusted with a lognormal distribution function, which can be further transformed to the number-weighted particle-size distribution shown in Fig. 4(d); for this transformation we assumed a spherical particle shape and used a value of $M_S = 301$ kA/m to relate the particle moments to the particle sizes. This distribution is in a good agreement with the size histogram determined with TEM, which in turn verifies the superparamagnetic magnetization behavior of MZFO-8. For the larger particles, the same approach (which assumes a Langevin-type magnetization behavior) results in size distributions that significantly deviate from the TEM results (data not shown). This is in line with the observed transition from superparamagnetic to ferromagnetic-like behavior with increasing size, similar to results reported in the literature [8,10,15].

## B. Unpolarized SANS measurements

We measured the total unpolarized SANS cross sections $d\Sigma/d\Omega$ of each sample at 10 different applied magnetic fields from 0 to 4 T at room temperature. Figure 5 (left panel) shows some selected two-dimensional SANS patterns (remanent state and 4 T), which contain nuclear and magnetic contributions. According to magnetometry, all the samples are nearly magnetically saturated at a field of 4 T [Fig. 4(a)]. Hence, the sector average of $d\Sigma/d\Omega$ parallel to the applied field ($\boldsymbol{q} // \boldsymbol{H}_0$) at 4 T is a good approximation to the purely nuclear SANS cross section $d\Sigma_{nuc}/d\Omega$ [compare also Eq. (1)]. As shown in Fig. 6, $d\Sigma_{nuc}/d\Omega$ in the high-$q$ range can be well described by a power law, $d\Sigma_{nuc}/d\Omega \propto q^{-4}$, which is expected in the Porod regime for orientationally-averaged particles with a discontinuous interface [37]. For both the MZFO-27 and MZFO-38 samples we observe peak structures in the scattering curves, which might be related to the narrow particle-size distribution [compare Fig. 2(b)]. By contrast, the MZFO-8 and MZFO-80 exhibit a relatively broad size distribution, which results in the absence of such features in the nuclear SANS.

Regarding the 2D patterns, Fig. 5 shows that the total (nuclear and magnetic) SANS cross sections $d\Sigma/d\Omega$ exhibit for all samples a weakly field-dependent (compare top panel in Fig. 7) and a nearly isotropic intensity distribution. This observation points towards the dominance of the isotropic nuclear scattering contribution. Since in general the nuclear SANS cross section is field independent, the magnetic SANS cross section $d\Sigma_M/d\Omega$ can be determined by subtracting, for each sample, the total $d\Sigma/d\Omega$ measured at the highest field of 4 T from the data at lower fields. The *field-dependent* $d\Sigma_M/d\Omega$ obtained in this way are displayed in Fig. 5 (right panel). It is seen that the intensity distributions of MZFO-8 and MZFO-80 are slightly anisotropic, elongated along the horizontal field direction, while the 2D $d\Sigma_M/d\Omega$ of MZFO-27 and MZFO-38 are isotropic. For MZFO-8 the angular anisotropy of $d\Sigma_M/d\Omega$ is found in a $q$-range that corresponds to an interparticle length scale, whereas MZFO-80 exhibits this anisotropy on an intraparticle length scale. This observation suggests for MZFO-80 the presence of transversal (perpendicular to $\boldsymbol{H}_0$) spin components, in line with the $|\widetilde{M}_y|^2 \cos^2(\theta)$ scattering contribution in Eq. (1).

The used procedure of subtracting the total unpolarized SANS scattering at a field close to saturation from data at lower fields (Figs. 5 and 7) suggests that it may not always be necessary to resort to polarization-analysis experiments in order to obtain the magnetic (spin-flip) SANS cross section. If nuclear-spin-dependent SANS and chiral scattering contributions are ignored, the comparison of the spin-flip SANS cross section (Eq. (18) in Ref. [18]) with the so-called spin-misalignment SANS cross section [obtained by subtracting from Eq. (1) the scattering at saturation $\propto |\widetilde{N}|^2 + |\widetilde{M}_z|^2 \sin^2(\theta)$] reveals that the subtraction procedure yields, except for the longitudinal magnetic term, a combination of (difference) Fourier components that is very similar to the spin-flip SANS cross section (albeit with different trigonometric weights). If the nuclear particle microstructure of the material under study does not change with the applied field (leaving aside magnetostriction effects), this procedure might be a practicable alternative to time-consuming and low-intensity polarization-analysis measurements.

The azimuthally averaged (over $2\pi$) $d\Sigma/d\Omega$ and $d\Sigma_M/d\Omega$ for each magnetic field value $H_0$ are summarized in Fig. 7. The magnitude of $d\Sigma_M/d\Omega$ is reduced compared to $d\Sigma/d\Omega$, which is due to the dominance of the nuclear scattering contributions in our systems. In the following, we will distinguish between the intraparticle ($q > q_c$) and the interparticle ($q < q_c$) $q$-ranges, which are roughly defined by the average particle sizes $D$ of the respective system (i.e., $q_c = 2\pi/D$). For each sample, $d\Sigma_M/d\Omega$ exhibits a strong and more pronounced magnetic field dependence as compared to $d\Sigma/d\Omega$ (Fig. 7).

Figure 8 displays the SANS results for the experimental intensity ratio $\alpha_{exp}$ as defined by Eq. (3). Dividing the $q$-range in regions corresponding to values larger or smaller than $q_c = 2\pi/D$, we can obtain information on either inter- or intraparticle moment correlations of the nanoparticles (we note that the high-$q$ range may also contain weak features due to interparticle correlations). Regarding the interparticle $q$-range ($q < q_c$), $\alpha_{exp}$ exhibits for all samples a strong $q$-dependence, which might be explained by a difference between the nuclear and magnetic structure factors [43]. However, *within* the intraparticle $q$-range, corresponding approximately to $q/q_c > 1$, we observe very distinct features. For the smallest nanoparticles (MZFO-8), $\alpha_{exp}$ is independent of $q$ and almost equals the theoretical limit given by Eq. (4). Based on the considerations of Sec. III. B, this then suggests a single-domain configuration of MZFO-8 with a homogeneous magnetization profile. For the case of nanoparticles with an intermediate diameter (MZFO-27 and MZFO-38), we observe a more or less pronounced peak in the intraparticle $q$-range, at $q \cong 0.34$ nm$^{-1}$ (MZFO-27) and at $q \cong 0.22$ nm$^{-1}$ (MZFO-38), while for the largest particles (MZFO-80) we observe a weak monotonic decrease of $\alpha_{exp}$ over the whole $q$-range. Similar peaks were reported in Ref. [35] and were attributed to inhomogeneous magnetization profiles. By increasing the applied magnetic field, the magnitude of the peak feature of the MZFO-38 sample decreases (Fig. 8 right panel), which strongly suggests the transition from an inhomogeneous to a homogenous spin structure, where the canted spins tend to align with respect to the magnetic field $\boldsymbol{H}_0$. As we will see below (Sec. IV.C), these observations are consistent with our micromagnetic simulations. In Fig. 8(a,b) the mere deviation from the horizontal line at large $q$-values may indicate the presence of an inhomogeneous internal spin structure of the larger nanoparticles.

To analyze in more detail the possible field-dependent transition from an inhomogeneous to a homogeneous spin structure for MZFO-38, we have extracted the corresponding pair-distance distribution functions $p(r)$ [Eq. (5)] from $d\Sigma_M/d\Omega$. We restricted our analysis to the intraparticle $q$-range, as visualized by the dashed vertical line in Fig. 9(a), and obtained the field-dependent $p(r)$ profiles shown in Fig. 9(b). Accordingly, these profiles approximately describe the scattering behavior in the intraparticle $q$-range. As can be seen in Fig. 9(b), at the highest field of 1.0 T the extracted distribution $p(r)$ is nearly bell-shaped [37], which indicates a homogeneous magnetization profile within the spherical nanoparticles, whereas with decreasing field the deviation of the profile from this ideal case increases. This feature is an additional strong indication for the transition from a homogenous to an inhomogeneous spin structure within the particle with decreasing field, and *vice versa*. We note that there exist many studies in the literature, employing other techniques such as Mössbauer spectroscopy, magnetic X-ray scattering, or photoemission electron microscopy, which also report an inhomogeneous nanoparticle spin structure and/or the presence of interparticle moment correlations (e.g. Ref [44–48]). To further support our experimental observations, we have performed numerical micromagnetic simulations of the size-dependent magnetization behavior of MZFO nanoparticle ensembles; these are discussed in the following.

### C. Micromagnetic simulations

In the micromagnetic simulations we have considered the four standard contributions to the total magnetic energy: energy in the external field, cubic magnetocrystalline anisotropy energy, and exchange and dipolar interaction energies. The nanoparticle microstructure, consisting of a distribution of Mn-Zn based nanoparticles, was generated by employing an algorithm described in Refs. [49–54]. The simulation volume (= sample volume) is a cubic box of size $\approx 300 \times 300 \times 300$ nm$^3$, which was discretized into $4 \times 10^5$ mesh elements with an average mesh size of 4 nm. The volume fraction of the nanoparticles was kept fixed at 80 %, leaving 20 % void. Materials parameters are: saturation magnetization $M_S = 480$ kA/m (typical for ferrites, see page 423 in Ref. [55]), anisotropy constant $K = 3 \times 10^3$ J/m$^3$ [56], and exchange-stiffness constant $A = 7 \times 10^{-12}$ J/m [53]. The equilibrium magnetization state of the system was found, as usual, by minimizing the total magnetic energy at a given value of the applied magnetic field. Periodic boundary conditions were applied in the simulations. For more details on our micromagnetic methodology, see Refs. [49–54].

Figure 10 depicts the sample microstructures used in the simulations. Since the sample volume is kept constant, an increase in the average particle size $D$ from 14 to 74 nm leads to a reduction of the particle number $N$, from $N \sim 40.000$ at 14 nm to $N \sim 40$ at 74 nm.

Figure 11 shows the field dependence of the quantity $|M|/M_S$ for different particle sizes. This parameter is defined as:

$$\frac{|M|}{M_S} = \frac{1}{N} \frac{\sum_{i=1}^{N}\left(M_{x,i}^2 + M_{y,i}^2 + M_{z,i}^2\right)^{1/2}}{M_S}, \quad (6)$$

which is a measure for the average deviation of the particle's magnetization state from the single-domain state, corresponding to $|M|/M_S = 1$. It becomes visible in Fig. 11 that (small) deviations from the uniform particle magnetization state appear for $D$-values ranging between 20-30 nm, which is in reasonable agreement with our conclusions from the SANS data analysis (compare Figs. 8 and 9). Since the micromagnetic algorithm does not take into account superparamagnetic fluctuations, the computed hysteresis curves in the inset of Fig. 11 cannot reproduce the experimentally observed transition from the superparamagnetic to the blocked regime [compare Fig. 4(a) and (b)]. It is seen that the quasi-static magnetization decreases with increasing particle size, since larger particles tend to be in a more nonuniform spin state than smaller particles. This is shown in Fig. 12, which displays the evolution of the parameter $|M|/M_S$ for each magnetic particle "$i$" and as a function of the applied field. Also shown are snapshots of the spin structure at selected fields, where the largest deviations from the uniform state are observed.

### V. CONCLUSION

In summary, the structure and magnetic properties of Mn-Zn ferrite (MZFO) single crystalline nanoparticles with average diameters ranging from 8 to 80 nm were investigated using a suite of experimental and simulation techniques. The increase of the remanent magnetization as well as the coercive field, determined from the magnetization curves, is a clear evidence for a transition from the superparamagnetic to the blocked state with increasing particle diameter. The analysis of the magnetic-field-dependent unpolarized SANS data demonstrates that the magnetization profiles of the larger nanoparticles deviate from the perfect single-domain state. This conclusion has mainly become possible by plotting a special intensity ratio (Eq. (3) and Fig. 8), originally introduced by Ernst, Schelten, and Schmatz [35]. Another important clue for the nonuniform internal spin structure was obtained by the computation of the pair-distance distribution function $p(r)$ (Fig. 9). The $p(r)$ data nicely confirm the field-dependent internal spin structure of the nanoparticles. In reasonable agreement with the outcome of the experimental data analysis, large-scale micromagnetic simulations reveal that slight deviations from single-domain behavior occur for Mn-Zn ferrite particle sizes above about 20-30 nm. In general, we emphasize that a fundamental understanding of magnetic SANS can only be obtained by comparing experimental data, both in Fourier and real space, to the results of simulations. The used procedure of subtracting the total unpolarized SANS scattering at or close to saturation from data at lower fields suggests that it may not always be necessary to perform challenging polarization-analysis experiments in order to obtain the magnetic SANS cross section. If the nuclear particle microstructure of the material under study does not change with the applied field, this procedure might be a practicable alternative to time-consuming and low-intensity polarized neutron measurements. Finally, we note that our study demonstrates the unique ability of SANS to detect even very small deviations of the magnetization configuration from the homogeneously magnetized state.

## ACKNOWLEGDEMENTS

The authors acknowledge the Heinz Maier-Leibnitz Zentrum for provision of neutron beamtime. We thank Jörg Schwarz and Jörg Schmauch (Universität des Saarlandes) for the technical support with the hydraulic press and for the SEM investigations. It is also a pleasure to thank Dirk Honecker for fruitful discussions. This research was supported by the EU-H2020 AMPHIBIAN Project (n. 720853). Philipp Bender and Andreas Michels thank the Fonds National de la Recherche of Luxembourg for financial support (CORE SANS4NCC grant).

## APPENDIX

### A. Materials

All the samples were prepared using commercially available reagents used as received. Benzyl ether (99 %), toluene (99 %), oleic acid (OA, 90 %), oleylamine (OAM, $\geq$ 98 %), manganese (II) acetylacetonate (Mn(acac)$_2$·2 H$_2$O $\geq$ 99 %), zinc chloride (ZnCl$_2$, $\geq$ 98 %), iron (III) chloride hexa-hydrate (FeCl$_3$·6 H$_2$O, 98 %), iron (II) chloride tetra-hydrate (FeCl$_2$·4 H$_2$O, 98 %), manganese chloride tetra-hydrate (MnCl$_2$·4 H$_2$O, $\geq$ 99 %), sodium hydroxide (NaOH, $\geq$ 98 %) were purchased from Aldrich Chemistry. Iron (III) acetylacetonate (Fe(acac)$_3$, 99 %) was obtained from Strem Chemicals and absolute ethanol (EtOH) was purchased from Fluka.

### B. Synthesis

The samples MZFO-27 and MZFO-38 were synthesized by thermal decomposition of iron and manganese acetylacetonates in high-boiling solvent (benzyl ether) in the presence of surfactants (OA, OAM) and of ZnCl$_2$. Instead, the samples MZFO-8 and MZFO-80 were prepared by the co-precipitation method using manganese chloride tetra-hydrate, zinc chloride, iron (II) chloride tetra-hydrate, iron(III) chloride hexa-hydrate, and sodium hydroxide as starting materials.

MZFO-27: Fe(acac)$_3$ (0.612 g, 1.733 mmol), Mn(acac)$_2$·2 H$_2$O (0.038 g, 0.133 mmol), ZnCl$_2$ (0.018 g, 0.133 mmol), OAM (2.675 g, 10 mmol), OA (2.825 g, 10 mmol) and benzyl ether (30 mL) ) were mixed and magnetically stirred under a flow of nitrogen in a 100 mL three-neck round-bottom flask for 15 min. The resulting mixture was heated to reflux (~ 290 °C) at 9 °C/min and kept at this temperature for 30 min under a blanket of nitrogen and vigorous stirring. The black-brown mixture was cooled to room temperature and EtOH (60 mL) was added causing the precipitation of a black material. The obtained product was separated with a permanent magnet, washed several times with ethanol, and finally re-dispersed in toluene.

MZFO-38: The synthesis and purification of this sample was carried out by following the same protocol used for MZFO-27, but using the metal/oleic acid/oleylamine ratio 1:5:5 and keeping the reaction mixture to reflux for 1 h.

MZFO-80: FeCl$_3$·6 H$_2$O (2.7 g, 10 mmol), FeCl$_2$·4 H$_2$O (0.597 g, 3 mmol), MnCl$_2$·4 H$_2$O (0.198 g, 1 mmol), ZnCl$_2$ (0.136 g, 1 mmol) and degassed water (10 ml) were mixed and magnetically stirred under a flow of nitrogen. The resulting mixture was added to a basic solution at 100 °C, obtained dissolving NaOH (1.72 g, 43 mmol) in degassed water (100 ml), and kept at this temperature for 2 h under a blanket of nitrogen and vigorous stirring. The black-brown mixture was cooled to room temperature and the obtained product was separated with a permanent magnet, washed several times with water, 2 times with ethanol and finally dried under nitrogen. The obtained product was annealed at 725 °C for 2 h under N$_2$ in a tubular furnace. The powder was finally mixed with OA and toluene, sonicated for 30', precipitated with a permanent magnet, and washed three times with ethanol.

MZFO-8: The synthesis and purification of this sample was carried out by following the same protocol used for MZFO-80, but using a larger amount of NaOH (2 g, 50 mmol) and OA (2 g, 7 mmol) as surfactant. This sample did not undergo any annealing step.

# FIGURE 1

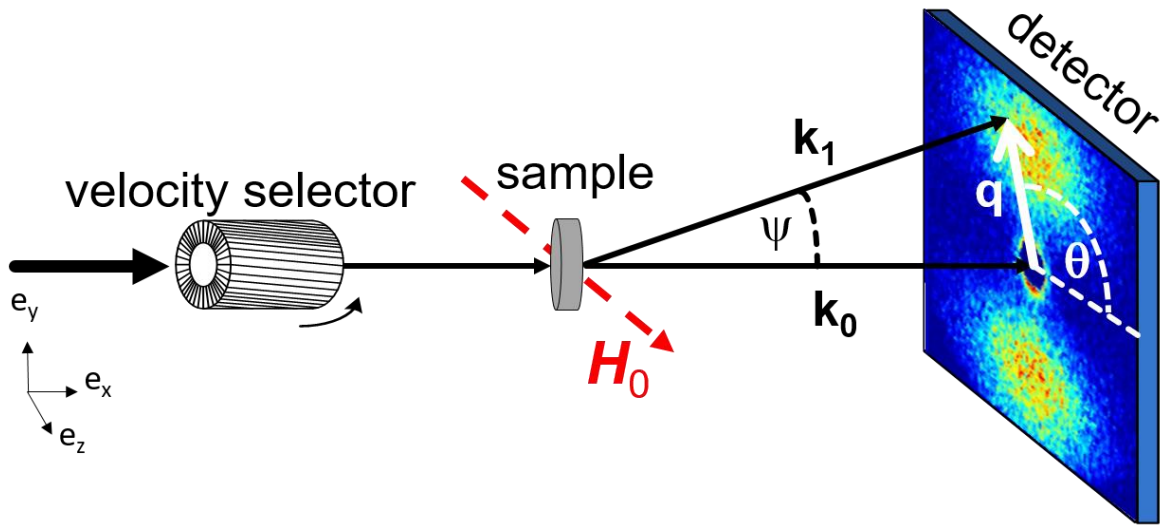

Fig. 1: Schematic drawing of the SANS setup. The scattering vector $q$ is defined as the difference between the wave vectors of the scattered and incident neutrons, i.e., $q = k_1 - k_0$. The magnetic field $H_0$ is applied perpendicular to the incident neutron beam, i.e., $k_0 \parallel \mathbf{e}_x \perp H_0 \parallel \mathbf{e}_z$. In the small-angle approximation ($\psi \ll 1$), the component of $q$ along $k_0$ is neglected, i.e., $q \cong \{0, q_y, q_z\} = q\{0, \sin(\theta), \cos(\theta)\}$, where the angle $\theta$ specifies the orientation of $q$ on the two-dimensional detector.

# FIGURE 2

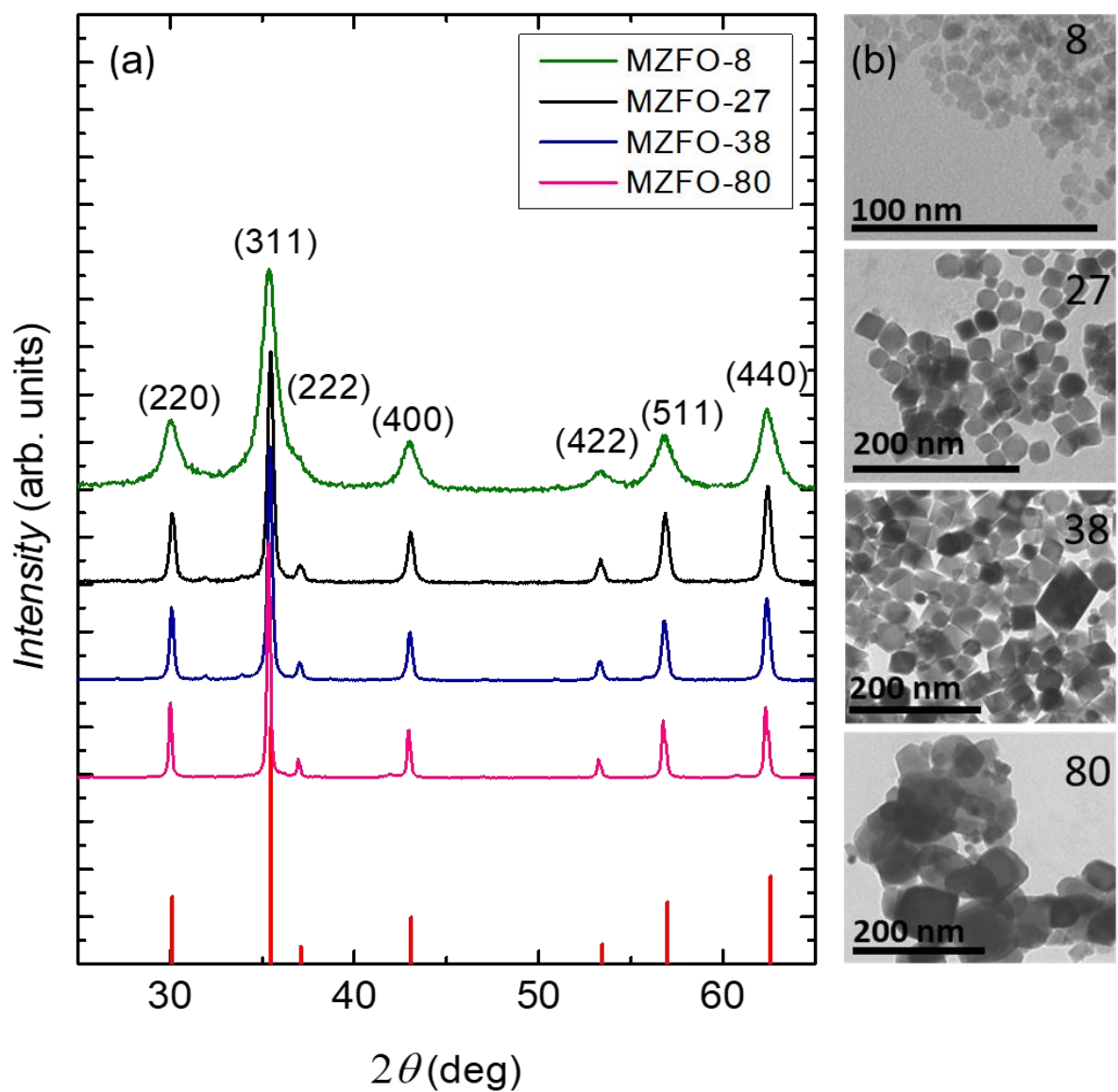

Fig. 2: (a) X-ray diffraction patterns of $Mn_{0.2}Zn_{0.2}Fe_{2.6}O_4$ nanoparticles (8, 27, 38, 80 nm diameter particle size) compared to the reference pattern of the cubic spinel structure (red color bars; taken from the JPCD database, JCPDS-221086). (b) TEM images of the $Mn_{0.2}Zn_{0.2}Fe_{2.6}O_4$ nanoparticles (8, 27, 38 and 80 nm diameter particle size).

# FIGURE 3

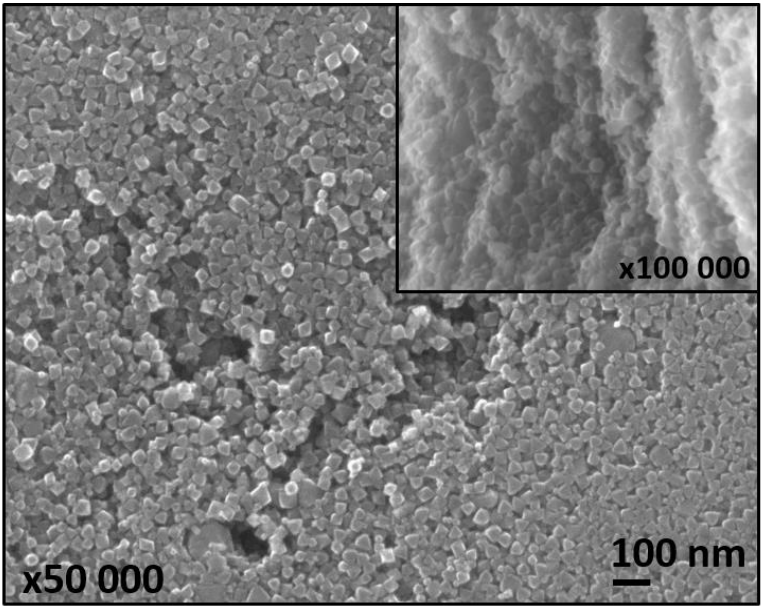

Fig. 3: Scanning electron microscopy (SEM) image of the MZFO-38 sample after pressing into a circular pellet. Inset: SEM cross view at the edge of the pellet.

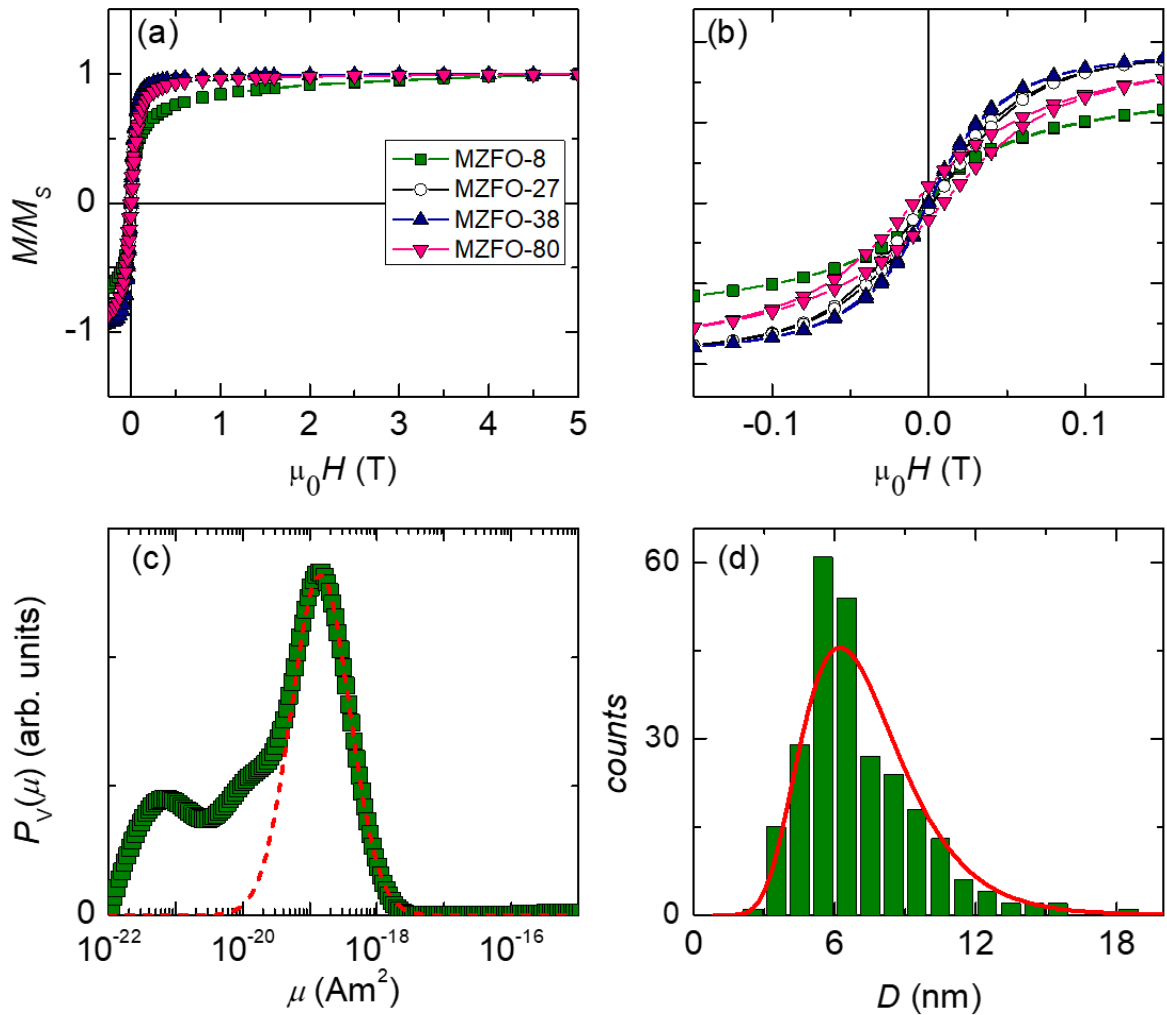

Fig. 4: (a) Normalized $M(H)$ curves of $Mn_{0.2}Zn_{0.2}Fe_{2.6}O_4$ nanoparticle powders measured at room temperature in a field range of ± 5 T (8 (green), 27 (black), 38 (blue) and 80 (pink) nm diameter particle size). The experimental $M_S$ has been approximated by the high-field value (5 T). (b) Zoom in the low-field region of the $M(H)$ curves. (c) Extracted magnetic moment distribution $P_V(\mu)$ of MZFO-8 determined by numerical inversion of the $M(H)$ in Fig. 4(a) (green squares). The main peak has been fitted assuming a log-normal distribution of the magnetic moment $\mu$ (red dashed line). (d) Histogram of the particle-size distribution of MZFO-8 determined by TEM (green) and number-weighted log-normal distribution determined by transforming the main peak of the magnetic moment distribution $P_V(\mu)$ observed in Fig. 4(c) (red solid line).

# FIGURE 5

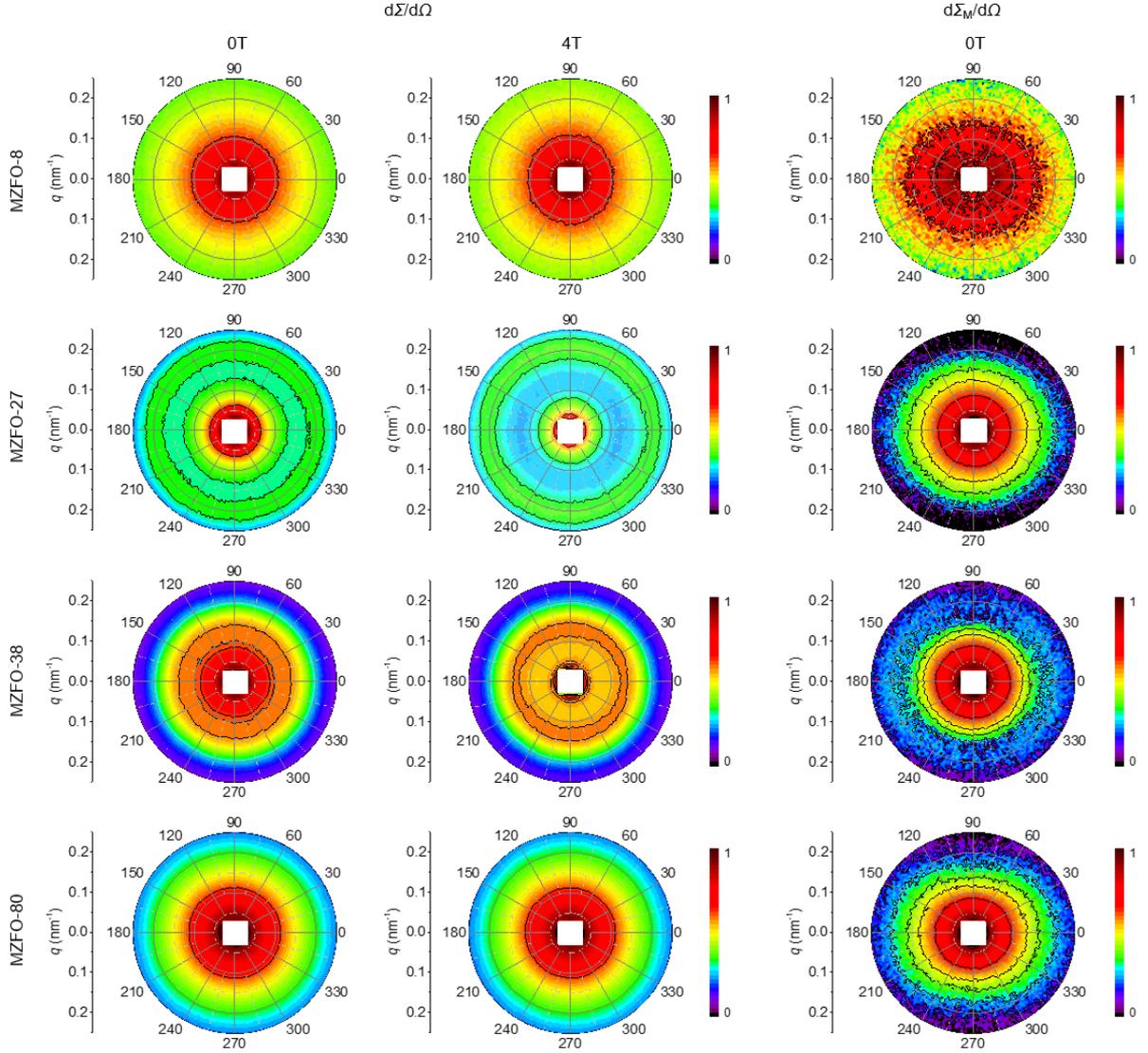

Fig. 5: Experimental two-dimensional total unpolarized SANS cross sections d$\Sigma$/d$\Omega$ (left and middle panel) and magnetic SANS cross sections d$\Sigma_M$/d$\Omega$ (right panel) of Mn$_{0.2}$Zn$_{0.2}$Fe$_{2.6}$O$_4$ nanoparticles. The d$\Sigma_M$/d$\Omega$ in the remanent state were obtained by subtracting the total scattering at the (near) saturation field of 4 T from the data at $H = 0$ T. The applied magnetic field $\boldsymbol{H}_0$ is horizontal in the plane of the detector ($\boldsymbol{H}_0 \perp \boldsymbol{k}_0$). All measurements were performed at room temperature. Note that the d$\Sigma$/d$\Omega$ and d$\Sigma_M$/d$\Omega$ scales are plotted in polar coordinates ($q$ in nm$^{-1}$, $\theta$ in degree and the intensity in arbitrary units normalized between 0 and 1).

# FIGURE 6

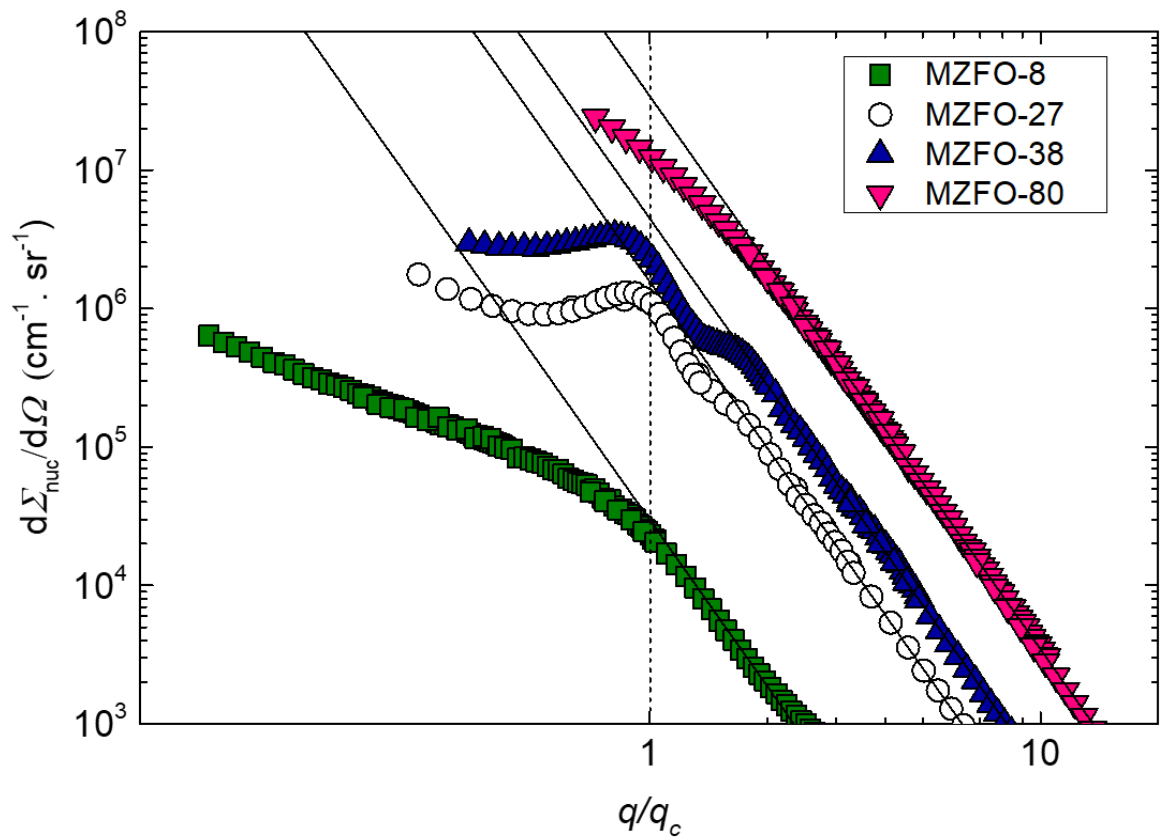

Fig. 6: Nuclear SANS cross sections $d\Sigma_{nuc}/d\Omega$ of $Mn_{0.2}Zn_{0.2}Fe_{2.6}O_4$ nanoparticles as a function of momentum transfer $q$ (8 (green), 27 (black), 38 (blue) and 80 (pink) nm diameter particle size) (log-log scale). The $d\Sigma_{nuc}/d\Omega$ were determined by ± 10° horizontal averages ($q // H_0$) of the total $d\Sigma/d\Omega$ at an applied magnetic field of $\mu_0H_0 = 4$ T. Note that the data are displayed as a function of $q/q_c$, where $q_c = 2\pi/D$ with $D$ the respective mean nanoparticle size. Measurements were performed at room temperature (300 K). Black solid lines: power law fits to $d\Sigma_{nuc}/d\Omega \propto K/(qD)^4$. Dashed vertical line: $q = q_c = 2\pi/D$. The error bars of $d\Sigma_{nuc}/d\Omega$ are smaller than the data point size.

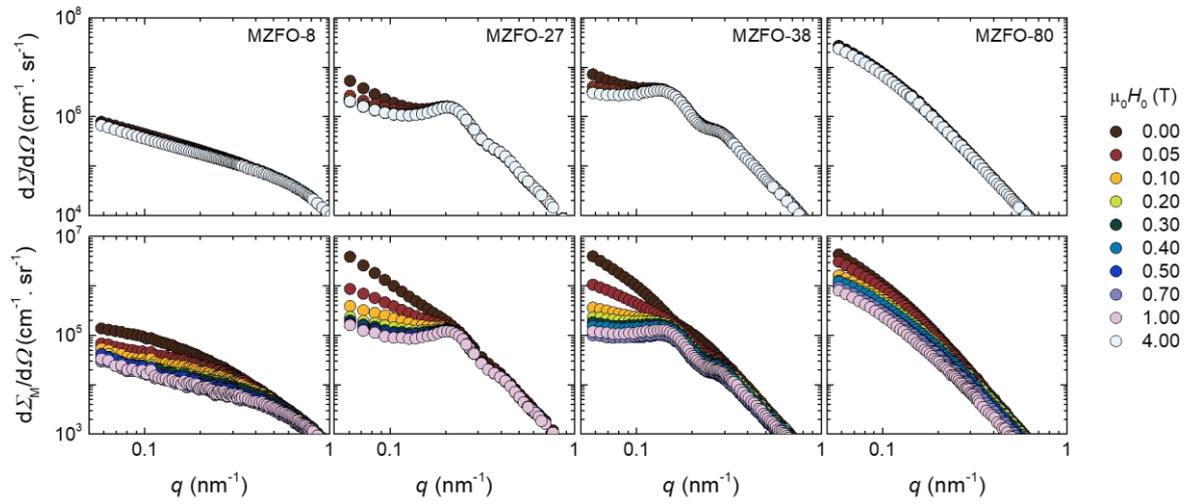

Fig. 7: Magnetic field dependence of the (over $2\pi$) azimuthally-averaged total nuclear and magnetic (top panel) and purely magnetic (bottom panel) SANS cross sections of $Mn_{0.2}Zn_{0.2}Fe_{2.6}O_4$ nanoparticles (log-log scale). Solid filled circles in the inset: magnetic field values in Tesla decrease from 4.0 T (bottom) to 0 T (top). All measurements were performed at room temperature. The error bars of $d\Sigma/d\Omega$ and $d\Sigma_M/d\Omega$ are smaller than the data point size.

# FIGURE 8

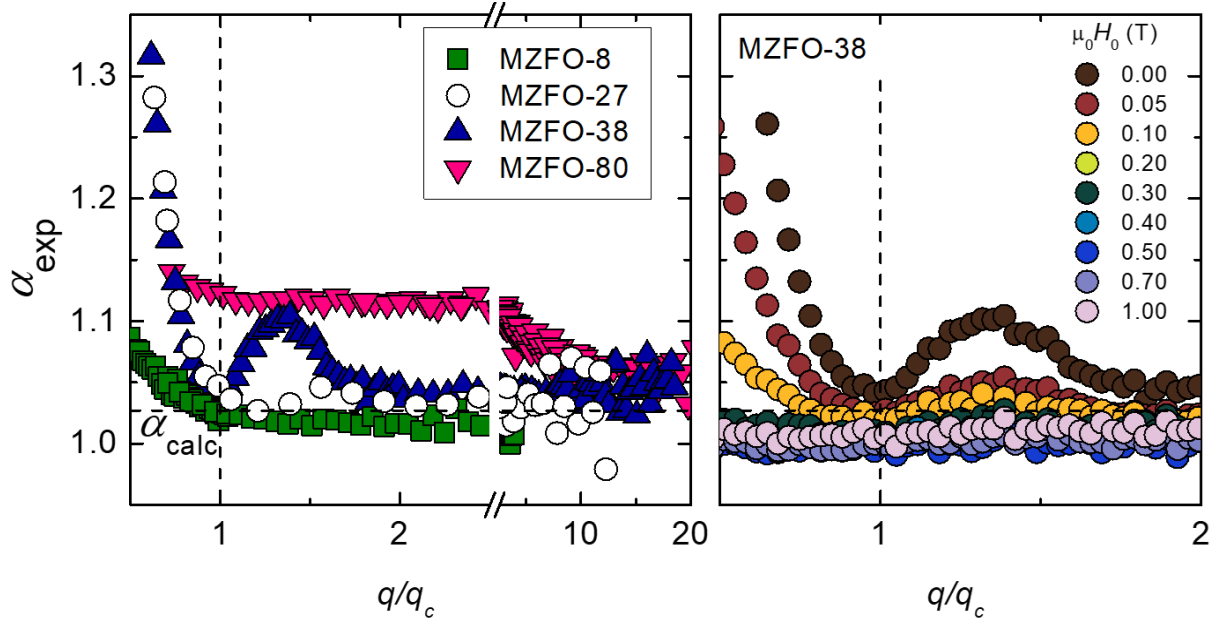

Fig. 8: Left: experimental intensity ratio $\alpha_{exp}(q)$ determined from the averaged SANS cross sections at zero field and at $\mu_0 H_0 = 4$ T with $\boldsymbol{q} \,//\, \boldsymbol{H}_0$ [Eq. (3)]. Right: magnetic field dependence of $\alpha_{exp}$ around $q = 0.22$ nm$^{-1}$ for MZFO-38. Note that the data are displayed as a function of $q/q_c$, where $q_c = 2\pi/D$ with $D$ the respective mean nanoparticle size. Dashed vertical lines: $q = q_c = 2\pi/D$. Dashed horizontal lines: $\alpha_{calc} = 1.027$ [Eq. (4)] computed using the Mn$_{0.2}$Zn$_{0.2}$Fe$_2$O$_4$ bulk density of 4084 kg/m³, $\Delta\varrho_{nuc} = \varrho_{Mn_{0.2}Zn_{0.2}Fe_2O_4} - \varrho_{Oleic\,acid} = 5.155\,\frac{10^{-6}}{\text{Å}^2}$, and $\varrho_{mag} = b_H M_S^{MZFO} = 1.046\,\frac{10^{-6}}{\text{Å}^2}$, where $M_S^{MZFO} = 359.4$ kA/m corresponds to the mean value of $M_S$ (compare Table 1).

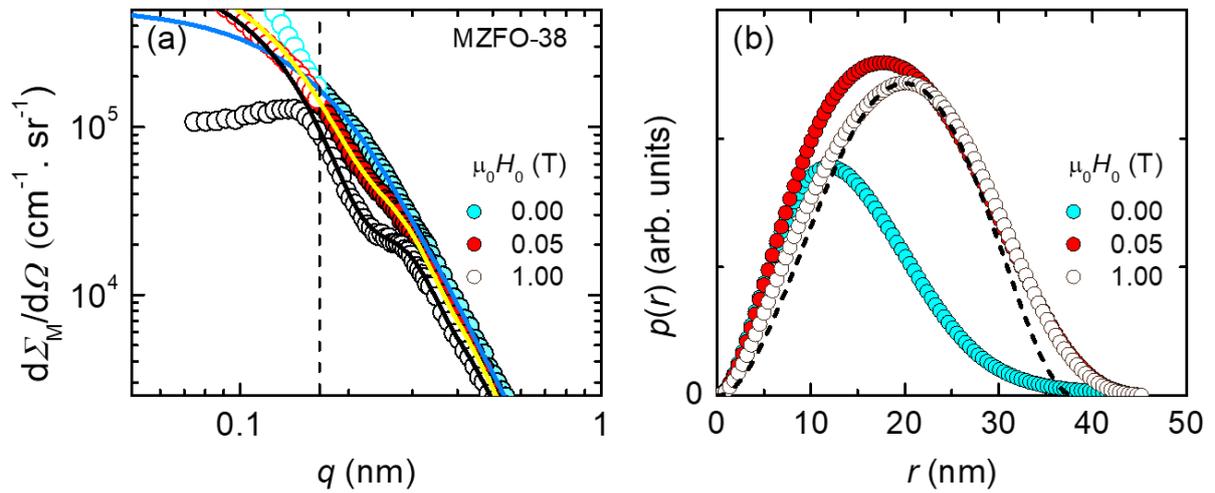

Fig. 9: (a) Selected field-dependent 2π-azimuthal averages of the magnetic SANS cross section $d\Sigma_M/d\Omega$ of MZFO-38 (taken from Fig. 7). Color solid lines: reconstruction of $d\Sigma_M/d\Omega$ in the intraparticle $q$-range (marked by the dashed vertical line) using the extracted $p(r)$ profiles from (b). (b) Field-dependent pair-distance distribution functions $p(r)$ [Eq. (5)] extracted by an indirect Fourier transform of $d\Sigma_M/d\Omega$ in the intraparticle $q$-range. Dashed line: expected $p(r) = r^2 [1 - 3r/(4R) + r^3/(16R^3)]$ for a homogeneous sphere of $D = 2R = 38$ nm size.

# FIGURE 10

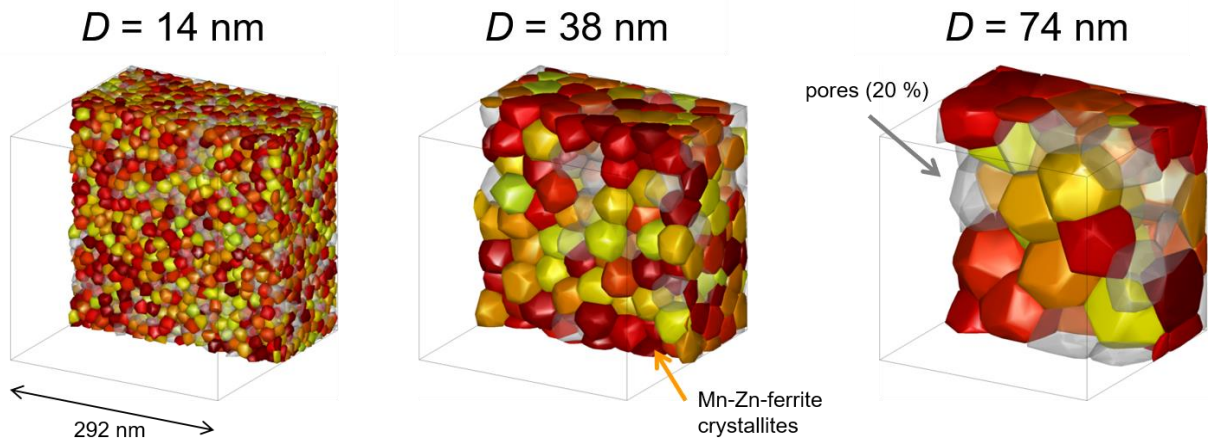

Fig. 10: Microstructures used in the micromagnetic simulations. The volume fraction of the particle phase was set to 80 % in all computations. The simulation volume ~ 300 × 300 × 300 nm$^3$ is constant in the simulations (mesh size: 4 nm), so that an increase in the average particle size $D$ is accompanied by a reduction of the number $N$ of particles, from $N$ ~ 40.000 at 14 nm to $N$ ~ 40 at 74 nm.

FIGURE 11

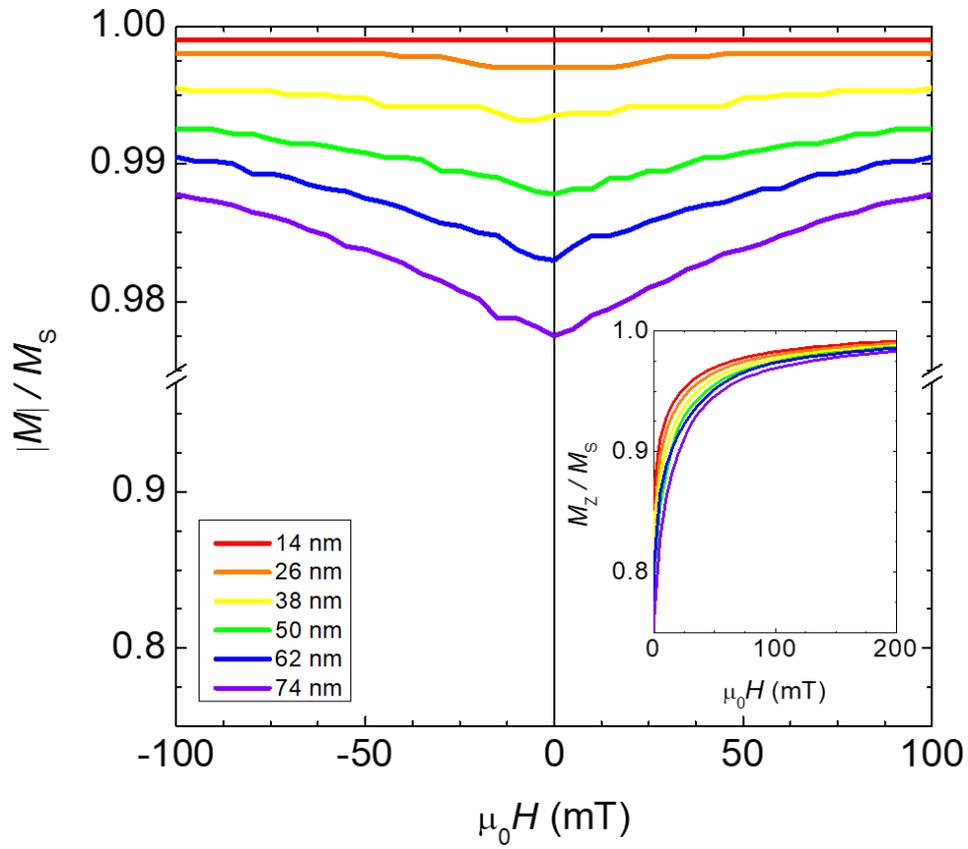

Fig. 11: Applied field dependence of the quantity $|M|/M_S$ [Eq. (6)] for different average particle sizes $D$. Inset: Corresponding normalized magnetization curves.

# FIGURE 12

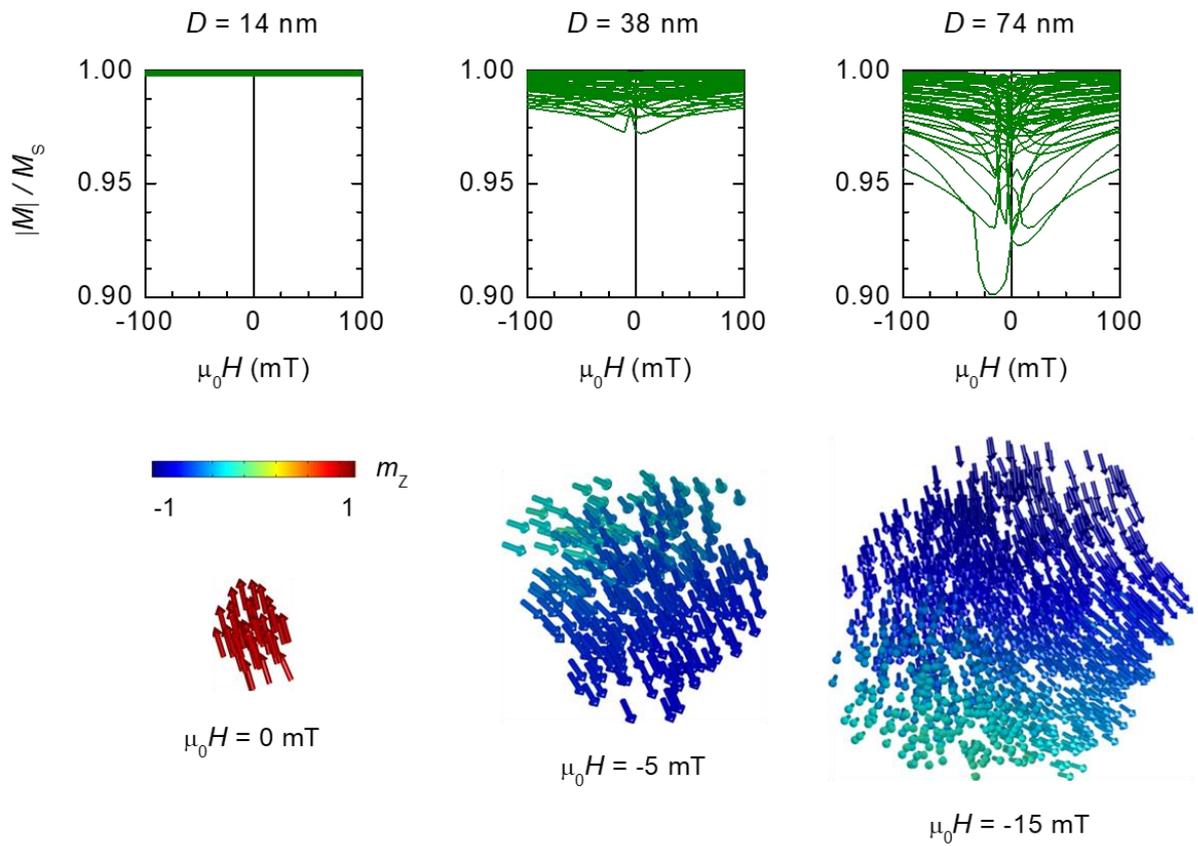

Fig. 12: (top panel) Particle-size-dependent evolution of the parameter $|M|/M_S$ [Eq. (6)] for each magnetic particle "$i$" and as a function of the applied magnetic field. (bottom panel) Snapshots of spin structures at selected fields, where the largest deviations from the uniform magnetization state are observed.

# TABLE 1

| Sample | Composition (XRF) | Particle size (TEM) (nm) | Crystal size (XRD) (nm) | $M_S$ at 300 K (Am$^2$/kg) | $M_R$ at 300 K (Am$^2$/kg) | $\mu_0 H_C$ at 300 K (mT) |
|---|---|---|---|---|---|---|
| MZFO-8 | $Mn_{0.18}Zn_{0.25}Fe_{2.57}O_4$ | 8±2 | 8(1) | 73 | 1 | 0.8 |
| MZFO-27 | $Mn_{0.24}Zn_{0.21}Fe_{2.55}O_4$ | 27±3 | 26(1) | 95 | 1 | 0.4 |
| MZFO-38 | $Mn_{0.20}Zn_{0.17}Fe_{2.63}O_4$ | 38±5 | 38(1) | 90 | 4 | 2.7 |
| MZFO-80 | $Mn_{0.20}Zn_{0.25}Fe_{2.55}O_4$ | ~ 80 | 79(1) | 94 | 10 | 9.4 |

Table 1: Structural and magnetic parameters of $Mn_{0.2}Zn_{0.2}Fe_{2.6}O_4$ nanoparticle powders. The average particle sizes were determined by means of transmission electron microscopy (TEM) and wide-angle X-ray diffraction (XRD). The saturation and remanent magnetizations ($M_S$ and $M_R$) and the coercive field ($H_C$) have been determined from the $M(H)$ curves.